\documentclass[12pt]{article}
\usepackage{cite}
\usepackage{amsmath,amsfonts,amssymb}
\usepackage{pstricks,pst-node}
\usepackage[small,bf,hang]{caption2}
\usepackage[dvips]{graphicx,epsfig}
\usepackage[vcentermath,enableskew]{youngtab}

\def\hybrid{
        \topmargin -20pt
        \oddsidemargin 0pt
        \headheight 0pt \headsep 0pt
        \textwidth 6.25in       
        \textheight 9.5in       
        \marginparwidth .875in
        \parskip 5pt plus 1pt   \jot = 1.5ex}

\hybrid
 \csname
@addtoreset\endcsname{equation}{section}

\def\moth{\mathsurround=0pt}
\newdimen\zo \zo=0pt

\def\tick{\leaders\hrule height 0.5ex depth 0pt \hskip 0.5pt}
\def\upboxfill{$\moth \setbox\zo\hbox{\tick}%
  \hskip 3pt\hbox to 0pt{$\tick$\hss}\hrulefill \hbox to 7.5pt{$\tick$\hss}$}

\def\dtick{\leaders\hrule height .34pt depth 0.5ex \hskip 0.5pt}
\def\downboxfill{$\moth \setbox\zo\hbox{\dtick}%
  \hskip 2pt\hbox to 0pt{$\dtick$\hss}\hrulefill \hbox to 2pt{$\dtick$\hss}$}

\def\bec{\begin{center}}
\def\ec{\end{center}}

\def\be{\begin{equation}}
\def\ee{\end{equation}}
\def\bea{\begin{eqnarray}}
\def\eea{\end{eqnarray}}
\def\ba{\begin{array}}
\def\ea{\end{array}}

\begin{document}

\begin{titlepage}
\begin{center}

\hfill MIT-CTP-4207 \\

\vskip 1.5cm

{\Large \bf T-duality versus Gauge Symmetry
\\[0.2cm]}

\vskip 1.5cm

{\bf Olaf Hohm\footnotemark}
\footnotetext{ohohm@mit.edu}

{\em Center for Theoretical Physics, Massachusetts Institute of Technology, \\
Cambridge, MA 02139, USA}
\vspace{.5cm}

\vskip 0.8cm

\end{center}

\vskip 1cm

\begin{center} {\bf ABSTRACT}\\[3ex]

\begin{minipage}{13cm}
\small{
We review the recently constructed `double field theory' which
introduces in addition to the conventional coordinates associated to momentum modes
coordinates associated to winding modes. Thereby, T-duality becomes a global symmetry of
the theory, which can be viewed as
an `$O(D,D)$ covariantization' of the low-energy effective space-time action of closed string theory.
We discuss its symmetries with a special emphasis on the relation between global
duality symmetries and local gauge symmetries. \\[0.5ex]
\textit{Note: This review article reports on work done in collaboration with Chris Hull and
Barton Zwiebach \cite{Hohm:2010jy,Hohm:2010pp} .}}

\end{minipage}

\vskip 3cm

\textit{Based on a talk presented at `String Field Theory and Related Aspects', \\
Kyoto, Japan, October 2010.}

\end{center}

\noindent

\vfill

January 2011

\end{titlepage}

\section{Introduction}
T-duality is perhaps the most intriguing duality of string theory, indicating a profound deviation
from our usual intuition about geometry. It relates circular or toroidal backgrounds $T^{D}$ with radius $R$
to backgrounds with radius $\alpha^{\prime}/R$ via the non-compact duality group $O(D,D)$ \cite{Giveon:1994fu}.
To fix ideas, let us start from the first-quantized string theory with world-sheet action
 \begin{equation}
   S \ = \ \int d^2\sigma \, E_{ij}(X)\,\partial_{+}X^{i}\partial_{-}X^{j}\,, \qquad
   E_{ij} \ = \ G_{ij}+B_{ij}\;,
  \end{equation}
where $G$ and $B$ are the background metric and B-field, respectively, and $\pm$ denote light-cone coordinates. If the background admits commuting isometries, i.e.,
if for a suitable choice of coordinate system $G$ and
$B$ are constant along coordinate directions corresponding to a torus,
the resulting two-dimensional field theory
is mapped to an equivalent theory under the transformation
  \begin{equation}\label{ODDtrans}
    E\,\rightarrow\, \left(aE+b\right)\left(cE+d\right)^{-1}\;, \qquad
     \begin{pmatrix} a &   b \\ c & d \end{pmatrix} \ \in \ O(D,D)\;,
  \end{equation}
which acts on the components of $E$ along the torus.
This `duality' is not an actual symmetry of the world-sheet
theory, for it involves a transformation of the background structure (i.e., of the `coupling constants')
rather than the physical fields.

The string theory description features in addition to momentum modes so-called winding modes,
and it is an old idea that the geometrical understanding of T-duality involves a corresponding doubling
of coordinates.
In other words, in addition to the usual momentum coordinates one has `winding-type' coordinates, and one
thinks of T-dualities as novel `coordinate transformations' that rotate momentum and winding coordinates
into each other.

Remarkably, in closed string field theory such a scenario is already realized \cite{Kugo:1992md}. 
Accordingly, based on this string field theory,  a space-time action can be determined 
\cite{Hull:2009mi}, at least perturbatively, in which the fields depend both on the usual
space-time coordinates $x^i$ and on winding coordinates $\tilde{x}_{i}$  (`double field theory').
Specifically, this action has been computed to cubic order around a flat background \cite{Hull:2009mi}. 
It reduces in the case that the
fields are independent of $\tilde{x}_{i}$ to the conventional low-energy
space-time action of closed string theory (expanded to cubic order in fluctuations),
 \begin{equation}\label{original}
  S \ = \ \int dx \sqrt{g}\,e^{-2\phi}\left[R+4(\partial\phi)^2-\frac{1}{12}H^2\right]\,,
 \end{equation}
where $\phi$ is the dilaton and $H=db$ the field strength of the Kalb-Ramond 2-form $b_{ij}$.

This double field theory action is background dependent, because the constant background
$E_{ij}=G_{ij}+B_{ij}$ enters explicitly. Consequently, $O(D,D)$ is not a proper symmetry,
but rather a duality: apart from a transformation of the fields, the invariance of the
action requires also an $O(D,D)$ transformation of the background $E_{ij}$ as in
(\ref{ODDtrans}). This feature is inherited from closed string field theory.

A background independent formulation of closed string theory is not known,
and therefore it is not clear how to formulate string theory in such a way that $O(D,D)$ becomes a
proper symmetry acting only on the dynamical variables.\footnote{In second-quantized closed string field theory,
despite not being background independent, there is actually a way to reformulate the $O(D,D)$ duality as a
symmetry. This is done by using classical solutions that connect backgrounds related by T-duality and which
modify the transformation rules of the string field by an inhomogeneous term \cite{Kugo:1992md}. 
}
More recently, however, a background independent formulation of
double field theory has been found \cite{Hohm:2010jy}, that can be viewed as a completion
to all orders for a certain subsector. To be more precise, we note that the cubic double field theory
requires the constraint $\tilde{\partial}^{i}\partial_{i}=0$, where
$\tilde{\partial}^{i}=\tfrac{\partial}{\partial\tilde{x}_{i}}$, which can be written in a
manifestly $O(D,D)$ covariant form,
 \begin{equation}\label{ODDconstr}
   \partial^{M}\partial_{M} \ = \ \eta^{MN}\partial_{M}\partial_{N} \ = \ 0\;, \qquad
   \   \eta^{MN} \ = \ \begin{pmatrix}
    0&1 \\1&0 \end{pmatrix}\,.
 \end{equation}
Here, $\eta$ is the $O(D,D)$ invariant metric, with indices $M,N,\ldots = 1,\ldots,2D$,
and $\partial_{M}=(\tilde{\partial}^{i},\partial_{i})$.
This constraint originates from the level-matching
condition in closed string theory, which here reduces to
$0=L_0-\bar{L}_0=-p_{i}w^{i}$, where $p^{i}$ are the momenta and $w_{i}$ the winding numbers.
The background independent double field theory action completes the construction for the subsector
that requires the stronger constraint that the differential operator (\ref{ODDconstr}) annihilates
not only all fields and gauge parameters, but also all of their products.

In this review, we introduce this background independent double field theory.
It admits a gauge symmetry that combines the usual diffeomorphisms and B-field
gauge transformations. While the usual gauge symmetries naturally give
rise to a `rigid' subgroup $GL(D,\mathbb{R})\ltimes\mathbb{R}^{D(D-1)/2}$
(in a sense which will be made precise below) this theory features
the larger semi-simple $O(D,D)$ as a global symmetry.
We discuss the relation between these global T-duality symmetries and
the local gauge symmetries, which in particular illuminates the physical consequences
of the strong form of the constraint (\ref{ODDconstr}).

\section{Double field theory with non-symmetric metric}
Since the T-duality group $O(D,D)$ acts on the sum of metric and B-field as in (\ref{ODDtrans}),
it is natural to assume that a background independent formulation involves the following
`non-symmetric' metric\footnote{Here we use a notation that does not distinguish between compact
and non-compact coordinates, and formally all coordinates are doubled. The case relevant for
string theory is the one where only compact coordinates are doubled, for which the fields depend
trivially on the new non-compact coordinates. Moreover, for a torus background the T-duality
group is actually $O(D,D;\mathbb{Z})$, but we will not always make explicit whether
$O(D,D)$ is a discrete or continuous symmetry.}
 \begin{equation}\label{calE}
  {\cal E}_{ij} \ = \ g_{ij}+b_{ij}\;,
 \end{equation}
combining the full metric and B-field, i.e., background plus fluctuation. The double field theory action
can indeed be written in terms of  (\ref{calE}) and reads
 \begin{equation}
\label{THEActionINTRO}
 \begin{split}\hskip-10pt
  S \ = \ \int \,dx d\tilde{x}~
  e^{-2d}\Big[&
  -\frac{1}{4} \,g^{ik}g^{jl}   \,   {\cal D}^{p}{\cal E}_{kl}\,
  {\cal D}_{p}{\cal E}_{ij}
  +\frac{1}{4}g^{kl} \bigl( {\cal D}^{j}{\cal E}_{ik}
  {\cal D}^{i}{\cal E}_{jl}  + \bar{\cal D}^{j}{\cal E}_{ki}\,
  \bar{\cal D}^{i}{\cal E}_{lj} \bigr)~
\\ &    + \bigl( {\cal D}^{i}\hskip-1.5pt d~\bar{\cal D}^{j}{\cal E}_{ij}
 +\bar{{\cal D}}^{i}\hskip-1.5pt d~{\cal D}^{j}{\cal E}_{ji}\bigr)
 +4{\cal D}^{i}\hskip-1.5pt d \,{\cal D}_{i}d ~\Big]\;.
 \end{split}
 \end{equation}
In here, $d$ is related to the dilaton via the field redefinition  $e^{-2d}=\sqrt{g}\,e^{-2\phi}$, and
the calligraphic derivatives are defined by
 \begin{equation}
  \label{groihffkdf}
  {\cal D}_i \ \equiv \ {\partial\over \partial x^i} - {\cal E}_{ik} \,{\partial \over \partial\tilde x_k}\,,
  ~~~~\bar {\cal D}_i \ \equiv \ {\partial\over \partial x^i} + {\cal E}_{ki}\, {\partial \over \partial\tilde x_k}\,.
 \end{equation}
This action is completely background independent and invariant under the following
generalization of the $O(D,D)$ action (\ref{ODDtrans}),
 \begin{equation}\label{ODDaction}
  {\cal E}^{\prime}(X^{\prime}) \ = \ (a{\cal E}(X)+b)(c{\cal E}(X)+d)^{-1}\;,
  \qquad
  d^{\prime}(X^{\prime}) \ = \ d(X)\;, 
 \end{equation}
where $X^{M}\equiv (\tilde{x}_{i},x^{i})$ combines the momentum and winding coordinates into a
fundamental $O(D,D)$ vector, and $X^{\prime}$ denotes the $O(D,D)$ rotated coordinates.
In (\ref{ODDaction}), metric and B-field are not assumed to admit any isometries, rather the
$O(D,D)$ action is well-defined on fields depending arbitrarily on the coordinates.  This symmetry is not manifest, but
nevertheless leaves each term in (\ref{THEActionINTRO}) separately invariant \cite{Hohm:2010jy}.

The action (\ref{THEActionINTRO}) admits a gauge symmetry with a parameter
$\xi^{M}\equiv (\tilde{\xi}_{i},\xi^{i})$ that combines the conventional
diffeomorphism and B-field gauge parameters $\xi^{i}$ and $\tilde{\xi}_{i}$, respectively,
\begin{equation}
 \label{finalgtINTRO}
 \begin{split}
  \delta {\cal E}_{ij} \ &= \ {\cal D}_i\tilde{\xi}_{j}-\bar{{\cal D}}_{j}\tilde{\xi}_{i}
  +\xi^{M}\partial_{M}{\cal E}_{ij}
  +{\cal D}_{i}\xi^{k}{\cal E}_{kj}+\bar{\cal D}_{j}\xi^{k}{\cal E}_{ik}\;,\\[0.5ex]
 \delta d \ &= \ \xi^M \partial_M d - {1\over 2}  \partial_M \xi^M\,.
 \end{split}
 \end{equation}
In order to verify the invariance of  (\ref{THEActionINTRO}) under (\ref{finalgtINTRO}),
the strong form of the constraint (\ref{ODDconstr}) is required, which is in contrast to the invariance
under $O(D,D)$.

This constraint is actually so strong that it implies that locally there is always an $O(D,D)$
transformation that rotates into a frame in which the fields depend only on `half of the
coordinates', say, the $x^{i}$. We next investigate the structure of this theory in different duality frames where
the fields either depend only on the momentum coordinates or only on the winding coordinates.

\subsection{Einstein gravity in different T-duality frames}
We consider the action (\ref{THEActionINTRO}) expanded in the number of winding-type
derivatives, $S = S^{(0)}+S^{(1)}+S^{(2)}$, where the superscript denotes
the number of $\tilde{\partial}$. The part $S^{(0)}$ is equivalent, up to a field redefinition, to the conventional
low-energy action (\ref{original}). The part $S^{(2)}$ quadratic in the winding derivatives is `T-dual'
to $S^{(0)}$. To see this, we consider
the transformation ${\cal E} \to \tilde {\cal E} = {\cal E}^{-1}$ which
is a special T-duality transformation (\ref{ODDaction}), corresponding to
an exchange of $x$ and $\tilde{x}$. $S^{(2)}$ can then be obtained from $S^{(0)}$ by
taking\cite{Hohm:2010jy}
 \begin{equation}\label{tildemapping}
  {\cal E}_{ij}\;\rightarrow\; \tilde{\cal E}^{ij}\;, \qquad
  g^{ij}\,\rightarrow\,\tilde{g}_{ij} \;, \qquad
  \partial_{i}\,\rightarrow\; \tilde{\partial}^{i}\;, \qquad
  d\;\rightarrow\; d\;.
 \end{equation}
Finally, $S^{(1)}$ is a mixed action with both momentum and winding derivatives that is needed
for gauge invariance.

It is instructive to inspect a similar expansion of the gauge transformations (\ref{finalgtINTRO}).
Writing out the calligraphic derivatives (\ref{groihffkdf}), we find
  \begin{eqnarray}\label{gaugeexpand}
   \delta_{\xi}{\cal E}_{ij} \,&=&\, {\cal L}_{\xi}{\cal E}_{ij}
   +\partial_{i}\tilde{\xi}_{j}-\partial_{j}\tilde{\xi}_{i} \\
   \nonumber
   &&+\,{\cal L}_{\tilde{\xi}}{\cal E}_{ij}-{\cal E}_{ik}
   \left(\tilde{\partial}^{k}\xi^{l}-\tilde{\partial}^{l}\xi^{k}\right){\cal E}_{lj}\;,
  \end{eqnarray}
where the first line involves only momentum derivatives and the second line only winding derivatives.
Here, ${\cal L}_{\xi}$ denotes the usual Lie derivative with respect to the parameter $\xi^{i}$.
Similarly, ${\cal L}_{\tilde{\xi}}$
denotes a Lie derivative in the winding coordinates, but now with respect to the
B-field gauge parameter $\tilde{\xi}_{i}$. Thus, we infer from (\ref{gaugeexpand}) that the
gauge transformations reduce for $\tilde{\partial}=0$ to the usual diffeomorphisms and
B-field gauge transformations. For non-vanishing winding
derivatives the gauge transformations become non-linear in the fields, as can be seen from the
second line in (\ref{gaugeexpand}). This has the curious consequence that the
gauge transformations mix metric and B-field. The first and second line are actually related by T-duality
in the same way as above: under the transformation  ${\cal E} \to \tilde {\cal E} = {\cal E}^{-1}$
together with $\xi^{i}\rightarrow \tilde{\xi}_{i}$ they are precisely interchanged.
In other words, what is a B-field gauge transformation in one duality frame becomes a
diffeomorphism in another duality frame, and vice versa.

The gauge transformations are actually covariant under arbitrary $O(D,D)$
transformations, but this is not manifest in the form (\ref{finalgtINTRO}). Next, we turn to a
reformulation of this double field theory involving the so-called generalized metric in
which the $O(D,D)$ covariance becomes manifest.

\section{Double field theory with generalized metric}
The generalized metric is a $2D\times 2D$ matrix that combines
metric and B-field in such a way that it transforms covariantly under $O(D,D)$,
 \begin{equation}\label{genmetric}
  {\cal H}^{MN} \ = \  \begin{pmatrix}    g_{ij}-b_{ik}g^{kl}b_{lj} & b_{ik}g^{kj}\\[0.5ex]
  -g^{ik}b_{kj} & g^{ij}\end{pmatrix}\;,
 \end{equation}
as indicated by its index structure. Remarkably, the double field theory action
(\ref{THEActionINTRO}) can be re-written in terms of ${\cal H}^{MN}$ such that
its $O(D,D)$ invariance becomes manifest \cite{Hohm:2010pp}, 
 \begin{equation}\label{Hactionx}
 \begin{split}
  S \ = \ \int dx d\tilde{x}\,e^{-2d}~\Big(~&\frac{1}{8}\,{\cal H}^{MN}\partial_{M}{\cal H}^{KL}
  \,\partial_{N}{\cal H}_{KL}-\frac{1}{2}{\cal H}^{MN}\partial_{N}{\cal H}^{KL}\,\partial_{L}
  {\cal H}_{MK}\\
  &-2\,\partial_{M}d\,\partial_{N}{\cal H}^{MN}+4{\cal H}^{MN}\,\partial_{M}d\,
  \partial_{N}d~\Big)\,.
 \end{split}
 \end{equation}

The gauge transformations that were non-linear in the form (\ref{finalgtINTRO})
become linear when written in terms of the generalized metric. They read
 \begin{equation}\label{manifestH}
  \delta_{\xi}{\cal H}^{MN} \ = \ \xi^{P}\partial_{P}{\cal H}^{MN}
  +(\partial^{M}\xi_{P} -\partial_{P}\xi^{M})\,{\cal H}^{PN}
  +
 ( \partial^{N}\xi_{P} -\partial_{P}\xi^{N})\,{\cal H}^{MP}\;,
 \end{equation}
where $O(D,D)$ indices are raised and lowered with $\eta^{MN}$.
This transformation rule is an $O(D,D)$ covariant extension of the
standard Lie derivative that governs infinitesimal diffeomorphisms,
with the new feature that each index gives rise to a `covariant' and
a `contravariant' contribution. In fact, in this language one can develop
a tensor calculus with generalized Lie derivatives that act on, say,
a `vector' according to
 \begin{equation}\label{genvec}
   \widehat{\cal L}_{\xi}A_{M} \ = \ \xi^{N}\partial_{N}A_{M}+
   \big(\partial_{M}\xi^{N}-\partial^{N}\xi_{M}\big)A_{N}\;,
 \end{equation}
and analogously on higher tensors with an arbitrary number of upper
and lower $O(D,D)$ indices. The gauge transformation (\ref{manifestH})
of the generalized metric then simply reduces to
$\delta_{\xi}{\cal H}^{MN}=\widehat{\cal L}_{\xi}{\cal H}^{MN}$.

Using these generalized Lie derivatives it is rather straightforward
to verify the closure of the gauge transformations, for it is sufficient
to check it on a vector $A_{M}$. One finds
  \begin{equation}
   \bigl[ \,  \widehat{{\cal L}}_{\xi_1}\,,  \widehat{{\cal L}}_{\xi_2} \, \bigr]\, A_M
   \ = \ - \widehat{{\cal L}}_{[\xi_1, \xi_2]_{{}_{\rm C}}}  A_M \,,
 \end{equation}
which defines a bracket (`C-bracket'),
\begin{equation}
 \label{cbracketdef}
  \bigl[ \xi_1,\xi_2\bigr]_{\rm{C}}^{M} \ \equiv
   \ \xi_{1}^{N}\partial_{N}\xi_{2}^M -\frac{1}{2}\,  \xi_{1}^P\partial^{M}\xi_{2\,P}
   -(1\leftrightarrow 2)\;.
 \end{equation}
This extends the Lie bracket characterizing the gauge algebra of the
usual diffeomorphisms. It is the double field theory extension of
the so-called Courant bracket in generalized geometry \cite{Hull:2009zb}. 
To see this we set $\tilde{\partial}=0$ in (\ref{cbracketdef}), which yields
for the vector and one-form component, respectively,
 \begin{equation}\label{cbr1}
  \big(\bigl[ \xi_1,\xi_2\bigl]_{\rm C}\big)^{i} \ = \ \xi_1^{j}\partial_{j}\xi_2^i
  -\xi_2^{j}\partial_{j}\xi_1^i \ \equiv \ \bigl[\xi_1,\xi_2\bigl]^{i}\;,
 \end{equation}
where $[\, ,\,]$ denotes the usual commutator or Lie bracket of vector fields, and
 \begin{equation}\label{cbr2}
 \begin{split}
  \big(\bigl[ \xi_1,\xi_2\bigl]_{\rm C}\big)_{i} \ &= \ \xi_1^{j}\partial_{j}\tilde{\xi}_{2 i}
  -\frac{1}{2}\xi^{j}_1\partial_{i}\tilde{\xi}_{2 j}-\frac{1}{2}\tilde{\xi}_{1 j}\partial_{i}\xi_2^{j}
  -(1\leftrightarrow 2)\\
  \ &= \ {\cal L}_{\xi_1}\xi_2-{\cal L}_{\xi_2}\xi_1 -\frac{1}{2}\partial_{i}
  \big(\tilde{\xi}_{2j}\xi_1^j\big)+\frac{1}{2}\partial_{i}
  \big(\tilde{\xi}_{1j}\xi_2^j\big)\;.
 \end{split}
 \end{equation}
In the mathematical literature on `generalized geometry' the central structure
is the direct sum of tangent and cotangent bundle over the base manifold $M$, i.e.,
$(T\oplus T^{\star})(M)$. The sections of this bundle can thus be viewed as formal
sums of vectors and one-forms, say $\xi+\tilde{\xi}$, where $\xi$ is the vector part
and $\tilde{\xi}$ is the one-form part. With this terminology, the results (\ref{cbr1})
and (\ref{cbr2}) can be summarized as
 \begin{equation}
  \bigl[ \xi_1+\tilde{\xi}_1,\xi_2+\tilde{\xi_2}\bigl] \ = \
  \bigl[ \xi_1,\xi_2\bigl] + {\cal L}_{\xi_1}\tilde{\xi}_2 - {\cal L}_{\xi_2}\tilde{\xi}_1
  -\frac{1}{2}d\big(i_{\tilde{\xi}_1}\xi_2-i_{\tilde{\xi}_2}\xi_1\big)\;,
 \end{equation}
where $i_{\tilde{\xi}_1}\xi_2\equiv \tilde{\xi}_{1i}\,\xi_2^{i}$, etc., denotes the
canonical product of a vector and a one-form.
This is precisely the so-called Courant bracket of generalized geometry \cite{Tcourant,Hitchin,Gualtieri}. 
The C-bracket can thus be seen
as the $O(D,D)$ covariant extension of the Courant bracket to which it reduces
in the `T-duality frame' $\tilde{\partial}^{i}=0$.

The gauge invariance of the double field theory, despite being non-manifest,
can be checked quite straightforwardly using the form of generalized Lie derivatives for
the gauge transformations, but the above actions can
also be brought into a more
`geometrical' form. More precisely, one can define a curvature scalar
${\cal R}$ that can be viewed as a function of $d$ and either
${\cal H}$ or ${\cal E}$, such that up to boundary terms
 \begin{equation}\label{masteractionINTRO}
  S \ = \ \int dxd\tilde{x}\,e^{-2d}\,{\cal R}({\cal E},d) \ = \  \int dxd\tilde{x}\,e^{-2d}\,{\cal R}({\cal H},d)\;.
 \end{equation}
In here, ${\cal R}$ transforms as a scalar and $e^{-2d}$ as a density,
 \begin{equation}\label{covtrans}
   \delta_{\xi}{\cal R} \ = \ \xi^{M}\partial_{M}{\cal R}\;, \qquad
   \delta_{\xi}\big(e^{-2d}\big) \ = \ \partial_{M}\big(\xi^{M}e^{-2d}\big)\;,
 \end{equation}
from which invariance of the action immediately follows.

This geometrical form can be related to a remarkable construction by
Siegel \cite{Siegel:1993th,Siegel:1993xq}. It is based on a frame field $e_{A}{}^{M}$
that is a generalized vector in the sense of (\ref{genvec}) and which carries a flat index $A$
corresponding to local $GL(D)\times GL(D)$ tangent space transformations.
This formalism features connections for this gauge symmetry and generalized
curvature tensors. The resulting curvature scalar ${\cal R}$ allows to define an invariant
action as in (\ref{masteractionINTRO}), which turns out to be equivalent to the double field
theory actions discussed here. The detailed
relation between the two formalisms is by now
well-understood \cite{Hohm:2010pp,Hohm:2010xe,Kwak:2010ew}.

\section{The relation between duality and gauge symmetry}
The double field theory features two symmetries: local gauge transformations parametrized by $\xi^{M}$,
extending the diffeomorphism and B-field gauge symmetries of the usual low-energy effective
action (\ref{original}), and global $O(D,D)$ T-duality transformations. Here, we will discuss the
relation between these two symmetries \cite{Hohm:2010jy}.

Conventional Einstein gravity has a global
$GL(D,\mathbb{R})$ symmetry, which is simply a subgroup of the
diffeomorphism group. Moreover, if the theory is coupled to an antisymmetric tensor $b_{ij}$
subject to the gauge symmetry $\delta b_{ij}=\partial_{i}\tilde{\xi}_{j}-\partial_{j}\tilde{\xi}_{i}$,
then there is an additional global shift symmetry. To be more precise,
we can choose the gauge parameters  to be
 \begin{equation}\label{Tparameter}
  \tilde{\xi}_{i} \ = \ -\frac{1}{2}e_{ij}x^{j}\;, \qquad
  \xi^{i} \ = \ x^{j}h_j^{~i}\;  ,
 \end{equation}
where $h$ is an arbitrary $D\times D$ matrix and $e$ is antisymmetric.
Insertion into the gauge transformations (given by the first line of (\ref{gaugeexpand})) then yields
the infinitesimal transformations
 \begin{equation}\label{gaugetransfff}
  \delta_{\xi}{\cal E}_{ij} \ = \
  \xi^{k}\partial_{k}{\cal E}_{ij}
  +\left({\cal E}\,h^{t}+h\,{\cal E}\right)_{ij}+e_{ij}\;.
 \end{equation}
These generate the group $GL(D,\mathbb{R})$,
parametrized by $h$, and $\mathbb{R}^{D(D-1)/2}$, acting as constant shifts $e_{ij}$.
Together, they constitute the group $GL(D,\mathbb{R})\ltimes\mathbb{R}^{D(D-1)/2}$.
This symmetry become particularly interesting in the context of Kaluza-Klein reduction on
a torus $T^{D}$, where it represents an independent rigid symmetry in addition to the lower-dimensional
diffeomorphisms.
In an actual Kaluza-Klein reduction of (\ref{original}) on a torus, however, one finds the full
semi-simple $O(D,D)$ as a rigid symmetry group. The remaining symmetry generators
that complete the so-called geometric subgroup $GL(D,\mathbb{R})\ltimes\mathbb{R}^{D(D-1)/2}$
into $O(D,D)$ have no interpretation in terms of the original
gauge symmetries of the higher-dimensional
theory, and are therefore sometime referred to as `hidden symmetries'.

In the double field theory we have realized the full $O(D,D)$ as a global symmetry prior to
any dimensional reduction. Thus, the immediate question arises whether the full $O(D,D)$
rather than only the geometric subgroup
can be seen as particular gauge symmetries parametrized by $\xi^{M}$.  This question will be
addressed in the following, which requires a careful inspection of the strong constraint
(\ref{ODDconstr}).

We start by considering infinitesimal $O(D,D)$ transformations by group elements of the form
${\bf 1} + T$, where $T$ takes values in the Lie algebra. For the Lie algebra we choose the basis
  \begin{eqnarray}\label{Liebasis}
  \left(\begin{array}{cc} h &
  0 \\ 0 & -h^{t} \end{array}\right)\;, \quad \left(\begin{array}{cc} 0 &
  e \\ 0 & 0 \end{array}\right)\;, \quad \left(\begin{array}{cc} 0 &
  0 \\ f & 0 \end{array}\right)\;,
 \end{eqnarray}
where, again, $h$ is an arbitrary  $D\times D$ matrix, while $e$ and $f$ are antisymmetric.
The corresponding $O(D,D)$ transformations (\ref{ODDaction}) are
  \begin{equation}
 \begin{split}
  h:&\qquad {\cal E}^{\prime}(X^{\prime}) \ = \ {\cal E}(X)+{\cal E}(X)\,h^{t}
  +h\,{\cal E}(X)\;, \\
  e:&\qquad {\cal E}^{\prime}(X^{\prime}) \ = \ {\cal E}(X)+e\;, \\
  f:&\qquad {\cal E}^{\prime}(X^{\prime})  \ = \ {\cal E}(X)-{\cal E}(X)\,f\,{\cal E}(X)\;.
 \end{split}
 \end{equation}
If we introduce an infinitesimal parameter according to $X^{\prime M} = X^{M}-\xi^{M}(X)$,
this can be written
in terms of variations $\delta {\cal E}(X) \equiv  {\cal E}'(X) - {\cal E}(X)$,
 \begin{eqnarray}\label{oddvar}
  \delta_{h}{\cal E} &=& \xi^{M}\partial_{M}{\cal E}+{\cal E}\,h^t+h\,{\cal E}\;, \\\label{oddvar1}
  \delta_{e}{\cal E} &=& \xi^{M}\partial_{M}{\cal E}+e\;, \\\label{oddvar2}
  \delta_{f}{\cal E} &=& \xi^{M}\partial_{M}{\cal E}-{\cal E}\, f\,{\cal E}\;,
 \end{eqnarray}
where we used matrix notation.

Next, we have to investigate whether there is a choice of gauge parameters that gives rise to
these transformations.
As we mentioned above, by virtue of the strong constraint locally one can always find a
T-duality frame where the fields depend only on the momentum coordinates $x^{i}$.
The constraint is then satisfied by all gauge parameters that also depends only on $x^{i}$.
We can thus choose the parameters as in (\ref{Tparameter}), which gives rise to
gauge transformations as in  (\ref{gaugetransfff}), in agreement with the $h$ and $e$ transformations
in (\ref{oddvar}) and (\ref{oddvar1}). Thus, as above, we see that the geometric subgroup emerges
as particular gauge transformations.

Now, the remaining transformations parametrized by $f$ (the `hidden symmetries') cannot be
realized in a similar way in conventional Einstein gravity, but in the double field theory
there are potentially more possibilities since one may choose
 \begin{equation}\label{fparameter}
  \tilde{\xi}_{i} \ = \ 0\;, \qquad
  \xi^{i} \ = \  -{\frac 1 2}f^{ij}\tilde{x}_{j}\;.
 \end{equation}
This turns out, however, to be in general inconsistent with the constraint.
In fact, this constraint requires $\partial^M \xi^i \,\partial_M A=0$
for all fields $A$, which implies with (\ref{fparameter})
\begin{equation}
\label{dfgsafads}
f^{ij}\partial _j A \ = \ 0 \;.
\end{equation}
As the fields will in general have an arbitrary dependence on $x^{i}$, this condition is
not satisfied. Therefore, (\ref{fparameter}) is not an allowed gauge parameter and $O(D,D)$
is in general not part of the gauge symmetries. However, in the context of Kaluza-Klein reduction
on a torus, the fields are assumed to be independent of certain coordinate directions,
and so in these directions (\ref{dfgsafads}) is satisfied. The gauge transformations 
(\ref{gaugeexpand}) then read
 \begin{equation}
  \delta_{\xi}{\cal E}_{ij} \ = \ -({\cal E}\,f\,{\cal E})_{ij}\;, 
 \end{equation}
which coincides with (\ref{oddvar2}).  Thus, in this case the
full $O(d,d)$ -- with $d$ denoting the internal dimension -- can be
interpreted as a remnant of the higher-dimensional gauge symmetries,
and therefore we achieved an interpretation of the `hidden symmetries.'

\section{Discussion}
The double field theory discussed here provides a space-time theory in which the T-duality
group $O(D,D)$ is realized as a symmetry prior to torus reductions or the presence of
isometries. This is achieved by virtue of doubling the coordinates. A crucial ingredient for
the consistency is the constraint (\ref{ODDconstr}). In its `weak' form this constraint is a
direct consequence of the level-matching condition in closed string theory, and it is very unlikely
that it can be relaxed. Moreover, even with this constraint, there would be solutions that depend
non-trivially both on $x$ and $\tilde{x}$. The background independent action discussed here,
however, requires the strong form of the
constraint, which implies that locally the fields depend only on half of the coordinates.
This theory exhibits nevertheless several intriguing features which, we believe, sheds light on the novel
geometrical structures expected in the complete theory.
The construction of the full double field theory subject only to the
weak constraint remains as an outstanding open problem.

We close with a few remarks on the physical meaning of the strongly constrained theory.
The space of solutions can be easily characterized: it consists of all solutions
of (\ref{original}) plus its $O(D,D)$ transformations. Put differently, for any solution of the strongly
constrained double field theory there exists an $O(D,D)$ transformation that rotates
it locally into a solution of (\ref{original}) depending only on $x^{i}$. Thus, one might be
tempted to conclude that this theory is a physically equivalent reformulation that makes
a certain symmetry manifest. This is not obvious, however. First, there might be a non-trivial
`patching' of coordinate charts such that, even though locally the fields depend either
only on the momentum coordinates or only on the winding coordinates, globally the solution is
non-trivial (`T-folds') \cite{Hull:2004in}. 
This possibility deserves further investigation.
Second, for the theory to be actually equivalent to the standard low-energy theory (\ref{original})
one would have to identify all solutions within a given $O(D,D)$ orbit. Let us illustrate this point
in a little more detail.

Suppose, we
start with an arbitrary solution $\{ g_{ij}(x),\,b_{ij}(x),\,d(x) \}$
of (\ref{original}), i.e., with fields depending only on the momentum coordinates.
If we act on this solution with an $O(D,D)$ element belonging to the geometric subgroup
$GL(D,\mathbb{R})\ltimes\mathbb{R}^{D(D-1)/2}$, then this transformation can be viewed
as a particular gauge transformation, as we discussed above. Thus, the two solutions simply represent
different parametrizations of the same physics and have to be identified. If we act
with a genuine T-duality transformation not belonging to the geometric subgroup
(corresponding to the `hidden symmetries' parametrized by $f$),
the situation is more subtle. In the case that the solution has an isometry along the direction we
T-dualize, the resulting transformation can still be viewed as a gauge transformation,
as we saw above.
Consequently, the resulting solution (still depending only on the $x^{i}$)
is physically equivalent to the original one. This is in agreement with our expectation that the
physics on toroidal backgrounds related by T-duality is the same.
If, on the other hand, the solution does not have an
isometry along the direction we are T-dualizing, then this transformation will finally switch on a dependence
on the winding coordinates $\tilde{x}_{i}$. Moreover, by the analysis of the previous section,
this is precisely the case for which the transformation \textit{cannot} be viewed as a
gauge symmetry. Thus, naively there is no reason to identify these two solutions, and so one would
conclude that the phase space of double field theory (viewed as the space of classical
solutions modulo gauge transformations) is larger.

This conclusion is, however, somewhat puzzling from the point of view that in closed string field theory,
which was the starting point for this construction but which requires only the weak constraint,
\textit{all} $O(D,D)$ transformations can be viewed as (discrete) gauge transformations (see also footnote above) \cite{Kugo:1992md}. 
Therefore, one would expect that if the double field theory discussed here can be generalized to a
`weakly constrained' theory, then all $O(D,D)$ transformations should become gauge transformations
and thus all solutions related by T-duality have to be identified. We conclude that
it remains as an open question what the precise physical meaning of this theory is
and how it will embed into the `ultimate' formulation of string theory.

\section*{Acknowledgements}
It is a pleasure to thank my collaborators Chris Hull and Barton Zwiebach. I would also like
to thank Seung Ki Kwak for collaboration on a related paper \cite{Hohm:2010xe}, 
Ashoke Sen, Martin Rocek and Stefan Vandoren for discussions, and  
the organizers of `String field Theory and Related Aspects' for 
providing a stimulating atmosphere. 

This work is supported by the U.S. Department of Energy (DoE) under the cooperative
research agreement DE-FG02-05ER41360 and by the DFG -- The German Science Foundation.


\begin{thebibliography}{99}


\bibitem{Giveon:1994fu}
  A.~Giveon, M.~Porrati and E.~Rabinovici,
  ``Target space duality in string theory,''
  Phys.\ Rept.\  {\bf 244}, 77 (1994)
  [arXiv:hep-th/9401139].

\bibitem{Kugo:1992md}
  T.~Kugo and B.~Zwiebach,
  ``Target space duality as a symmetry of string field theory,''
  Prog.\ Theor.\ Phys.\  {\bf 87} (1992) 801
  [arXiv:hep-th/9201040].

\bibitem{Hull:2009mi}
  C.~Hull and B.~Zwiebach,
  ``Double Field Theory,''
  JHEP {\bf 0909} (2009) 099
  [arXiv:0904.4664 [hep-th]].

\bibitem{Hohm:2010jy}
  O.~Hohm, C.~Hull and B.~Zwiebach,
  ``Background independent action for double field theory,''
  JHEP {\bf 1007} (2010) 016
  [arXiv:1003.5027 [hep-th]].

\bibitem{Hohm:2010pp}
  O.~Hohm, C.~Hull and B.~Zwiebach,
  ``Generalized metric formulation of double field theory,''
  JHEP {\bf 1008} (2010) 008
  [arXiv:1006.4823 [hep-th]].

\bibitem{Hull:2009zb}
  C.~Hull and B.~Zwiebach,
  ``The gauge algebra of double field theory and Courant brackets,''
  JHEP {\bf 0909} (2009) 090
  [arXiv:0908.1792 [hep-th]].



 \bibitem{Tcourant}
 T.~Courant, ``Dirac Manifolds."  Trans. Amer. Math. Soc. {\bf 319}: 631-661, 1990.

  %
\bibitem{Hitchin}
N.~Hitchin,
``Generalized Calabi-Yau manifolds,''
 Q. J. Math.  {\bf 54}  (2003), no. 3, 281--308,
arXiv:math.DG/0209099.
%

\bibitem{Gualtieri}
M.~Gualtieri,
``Generalized complex geometry,"
PhD Thesis (2004).
arXiv:math/0401221v1 [math.DG]


\bibitem{Siegel:1993th}
  W.~Siegel,
  ``Superspace duality in low-energy superstrings,''
  Phys.\ Rev.\  D {\bf 48}, 2826 (1993)
  [arXiv:hep-th/9305073].


\bibitem{Siegel:1993xq}
W.~Siegel,
  ``Two vierbein formalism for string inspired axionic gravity,''
  Phys.\ Rev.\  D {\bf 47}, 5453 (1993)
  [arXiv:hep-th/9302036].

\bibitem{Hohm:2010xe}
  O.~Hohm and S.~K.~Kwak,
  ``Frame-like Geometry of Double Field Theory,''
  arXiv:1011.4101 [hep-th].

\bibitem{Kwak:2010ew}
  S.~K.~Kwak,
  ``Invariances and Equations of Motion in Double Field Theory,''
  JHEP {\bf 1010} (2010) 047
  [arXiv:1008.2746 [hep-th]].

\bibitem{Hull:2004in}
  C.~M.~Hull,
  ``A geometry for non-geometric string backgrounds,''
  JHEP {\bf 0510} (2005) 065
  [arXiv:hep-th/0406102].


\end{thebibliography}
\end{document}